# On the method of ascertainment
# of the possible historical proximity of two Extensive Air Showers

Yuri Verbetsky, Manana Svanidze

E Andronikashvili Institute of Physics under Tbilisi State University, Tbilisi, Georgia
Email:   yuverbetsky@mail.ru;   mananasvanidze@yahoo.com

***Abstract***:  *The increasing efforts are still in progress to establish existence and to investigate the properties of pairs of Extensive Air Showers (EAS) that can be considered as originated from a single event which produced the Cosmic Radiation (CR) particles that in turn initiated both showers. Considerably remote CR installations observing EAS events are particularly useful for this purpose. The estimation method is proposed for determination of such EAS pairs, observed by a pair of mutually remote installations and initiated by two initial CR particles, which can be regarded as historically proximal and possibly congenetic. The numerical example of application of this method is given using a toy simulation sample of showers.*
***Keywords***: *primary cosmic radiation, extensive air shower, couple of showers, relativistic invariant parameters, proximity determination.*

## Introduction

Investigations of properties of Primary Cosmic Radiation (PCR) have led to the need of research of properties of pairs of remote Extensive Air Showers (EAS or simply "showers") generated by the PCR particles in the Earth atmosphere and observed by mutually spaced installations. This new requirement is caused by the fact that some fraction of these EAS pairs (not necessarily the showers of very high energy) can be produced by PCR particles originated in some single interaction in the space. That is why these historically related pairs of PCR particles (the EAS ancestors) and consequently the induced showers can yield additional information on properties of PCR, such as e.g. the PCR chemical composition, interplanetary and interstellar medium properties (e.g. the medium density distribution), and can possibly provide insight into the Dark Matter properties [1]. However, only a few shower-pair examples are known [2 - 7] so far. This does not allow determination of phenomenological properties of the flux of such pairs, let alone the properties of their source.

So long as the observation rate of the historically related EAS pairs is low the big total area of the remote Cosmic Ray (CR) stations for the detection of EAS events is necessary to investigate the problem. That is why some widespread national networks of small-scale, simple and inexpensive CR stations are executing the respective investigations (e.g. ALTA, CZELTA, HiSPARC, LAAS, Showers of Knowledge, EEE, etc.). A new international association investigating the problem, i.e. Cosmic-Ray Extremely Distributed Observatory (CREDO [8]), has been organized recently. The network GELATICA [9 - 12] in Georgia has acceded to this association along with some of the national networks. Some of the CR stations of these networks have a capability to detect the arrival direction of the shower in addition to the shower's occurrence time. The ground-based installations with such capabilities hereinafter are referred to as "EAS goniometers" or simply "goniometers".

At the current initial stage of the investigation one needs to establish a selection method for EAS pairs initiated by the PCR particles, possibly spatially proximal in their past. In this case the separated shower pairs can be considered as the desired congenetic EAS couples.

For this purpose the special relativistic invariant kinematic quantities are designed on the basis of available data from two EAS motion characteristics estimated by two remote goniometers. These quantities can be used to ascertain the historical proximity of the ancestors of both showers in the pair. The method of the verifying parameters design is proposed for separation of the needed congenetic EAS couples from the whole set of the observed EAS random pairs.

*Determination of the historical proximity measure of two extensive air showers*

It is essential to establish the measure of the historical proximity of two showers as the minimal distance ever separating the two ancestor CR particles. However, this proximity must be interpreted with an allowance for estimation error.

Let us accept the laboratory system of coordinates with fixed radius-vectors $\mathbf{r}_{01}$, $\mathbf{r}_{02}$ of two remote shower cores' observation. This lab system, common for both goniometers, is necessarily equivalent to the geocentric coordinate system. Let us use the universal time UTC $\hat{t}$ (common for the Earth globe) as the laboratory time. Since the gravitational effects are negligible, the lab system can be considered as the inertial one.

For the purpose of the current special problem of determination of two-shower historical proximity measure it is sufficient to consider any EAS as a free-moving point within the measurement accuracy with the region of the shower's charged-particle greatest density, i.e. with the shower's core. Let us assume that this shower-connected representing point is permanently moving along a straight line with the velocity $c$ of light in free space. This assumption is acceptable for the CR particle with the energy sufficient for EAS origination in the Earth atmosphere since the true atmospheric track of the EAS core is almost straight and the ancestor particle is practically undeflected by the surrounding electric and magnetic fields at such energy. The motion equation of the mentioned representing point of the shower and its ancestor particle is a familiar equation of uniform motion under the accepted approximation.

The physical system under study consists of two separate showers observed by two remote goniometers at two different lab-times $\hat{t}_{01}$, $\hat{t}_{02}$ with the cores detected at remote points $\mathbf{r}_{01}$, $\mathbf{r}_{02}$ and the motion directions described by two *unit dimensionless* vectors (two "orts") $\mathbf{n}_1$, $\mathbf{n}_2$ of motion velocity. The rare occurrences with parallel motion ($\mathbf{n}_1 = \mathbf{n}_2$) of both showers are excluded. The available initial data from two goniometers for two observed showers are specified in the table 1 together with the respective dispersions and covariance matrices of their estimation.

**TABLE 1    INITIAL DATA**

| The estimated initial values | Designations |
|---|---|
| Radius-vectors of two observation points of the shower cores; the respective covariance matrices | $\mathbf{r}_{01}$, $\mathbf{r}_{02}$ $\mathbf{M}_1$, $\mathbf{M}_2$ |
| The universal times (UTC) of those shower cores' observations; the respective dispersions | $\hat{t}_{01}$, $\hat{t}_{02}$ $\sigma_{t1}^2$, $\sigma_{t2}^2$ |
| Orts of motion directions of two showers; the respective covariance matrices | $\mathbf{n}_1$, $\mathbf{n}_2$ $\mathbf{D}_1$, $\mathbf{D}_2$ |

Certainly, the coordinates of core detections may not by available for any goniometer. Though, one may substitute these data by the proper coordinates of the goniometers themselves, under the condition of sufficiently big spatial separation of installations with respect to the root-mean-square distance from the facility to the cores of the observable showers. The latter RMS distance can be used as the uncertainty measure of the coordinates of the shower cores.

The usual equations of uniform motion of both representing points along two skew straight lines are $\mathbf{r}_1(\hat{t}) = \mathbf{r}_{01} + c(\hat{t} - \hat{t}_{01}) \cdot \mathbf{n}_1$ and $\mathbf{r}_2(\hat{t}) = \mathbf{r}_{02} + c(\hat{t} - \hat{t}_{02}) \cdot \mathbf{n}_1$. Therefore the variable vector connecting these two moving representing points *at a time instant* $\hat{t}$ depends linearly on $\hat{t}$:

$$\tilde{\Delta}(\hat{t}) = \mathbf{r}_2(\hat{t}) - \mathbf{r}_1(\hat{t}) = \underbrace{\left[(\mathbf{r}_{02} - \mathbf{r}_{01}) - c(\hat{t}_{02} \cdot \mathbf{n}_2 - \hat{t}_{01} \cdot \mathbf{n}_1)\right]}_{\text{const}} + [c(\mathbf{n}_2 - \mathbf{n}_1)] \cdot \hat{t} \qquad (1)$$

It is reasonable to use the special zero-time reference in the following treatment of every shower pair as a separate system. Let us take the average time $\hat{t}_c = (\hat{t}_{02} + \hat{t}_{01})/2$ of two observation times (UTC) of both showers for the system time origin. This choice helps to avoid unreasonably large numerical values in the following calculations. Thus *the system time* of any event is defined as $t = \hat{t} - \hat{t}_c$. All the times values within the given system of two showers are *the coordinate system times* hereinafter.

The showers' observation times under the system time origin are

$$t_1 = \hat{t}_{01} - \hat{t}_c = -\delta t/2; \qquad t_2 = \hat{t}_{02} - \hat{t}_c = \delta t/2; \qquad (2)$$

Here the time difference $\delta t = \hat{t}_{02} - \hat{t}_{01}$ can have any sign, so the shower observation times $t_1$, $t_2$ are equal-in-magnitude and opposite in sign. The earliest observation occurs at $-|\delta t|/2$ moment, at negative system time.

Let us designate by $\delta \mathbf{r} = (\mathbf{r}_2 - \mathbf{r}_1)$ the vector connecting both points of the shower cores' observation; by $\delta \mathbf{n} = (\mathbf{n}_2 - \mathbf{n}_1)$ the difference vector of the showers' velocity orts; and by $\langle \mathbf{n} \rangle = (\mathbf{n}_1 + \mathbf{n}_2)/2$ the average vector of those orts. The last two vectors are mutually orthogonal: $(\langle \mathbf{n} \rangle^T \cdot \delta \mathbf{n}) = 0$. [The symbol $^T$ indicates the transpose of an algebraic vector or matrix].

The simple secondary values derived on the basis of initial data shown in the <u>table 1</u> are specified in <u>table 2</u> for referencing purpose. They are used in the following calculations.

TABLE 2      SIMPLE SECONDARY VALUES

| Derived values | Definitions |
|---|---|
| Time difference of the showers' observations; <br> the respective dispersion | $\delta t = \hat{t}_{02} - \hat{t}_{01}$ <br> $\sigma_{\delta t}^2 = \sigma_{t1}^2 + \sigma_{t2}^2$ |
| The showers' observation times; <br> the respective dispersions | $t_1 = -\delta t/2;\ t_2 = \delta t/2;$ <br> $\sigma_{\delta t}^2/4$ |
| The vector connecting both points of the shower cores' observation; <br> the respective covariance matrix | $\delta \mathbf{r} = \mathbf{r}_{02} - \mathbf{r}_{01}$ <br> $\mathbf{M} = \mathbf{M}_1 + \mathbf{M}_2$ |
| The difference vector of the showers' velocity orts; <br> the respective covariance matrix | $\delta \mathbf{n} = \mathbf{n}_2 - \mathbf{n}_1$ <br> $\mathbf{D} = \mathbf{D}_1 + \mathbf{D}_2$ |
| The average vector of the showers' velocity orts; <br> the respective covariance matrix | $\langle \mathbf{n} \rangle = (\mathbf{n}_1 + \mathbf{n}_2)/2$ <br> $(\mathbf{D}_1 + \mathbf{D}_2)/2^2 = \mathbf{D}/4$ |

The expression <u>(1)</u> for the variable vector connecting two moving representing points has simpler form when using the above definitions:

$$\tilde{\Delta}(t) = [\delta \mathbf{r} - \langle \mathbf{n} \rangle (c\, \delta t)] + \delta \mathbf{n} \cdot ct \qquad (3)$$

The smallest value of this vector's length (in the whole unbounded period of the showers' representing points' motion) will be assumed as the source for design of the desired measure of the showers' historical proximity. Let us use for this propose the relativistic invariant quantity, the squared interval between the representing points at any time instance $t$. The spatial vector $\tilde{\Delta}(t)$ connecting the two events (3) is known. The time difference between them is zero by definition. So the squared interval is $s^2(t) = -(c \cdot 0)^2 + \tilde{\Delta}(t)^T \cdot \tilde{\Delta}(t)$. This function has the smallest value at the moment $\tau$:

$$\tau = -\frac{1}{c} \cdot \frac{(\delta \mathbf{r}^T \delta \mathbf{n})}{(\delta \mathbf{n}^T \delta \mathbf{n})};$$

For $\tau < 0$, i.e. in the system past, the representing points have been in the nearest position. For $\tau > 0$ the representing points seem approaching each other, although in reality they are absorbed in the underlying ground.

The considered scheme of simultaneous motion of two representing points along two straight skew lines is shown in the figure 1. The special case of the points' closest approach *before* the earliest observation of one of the showers is shown here.

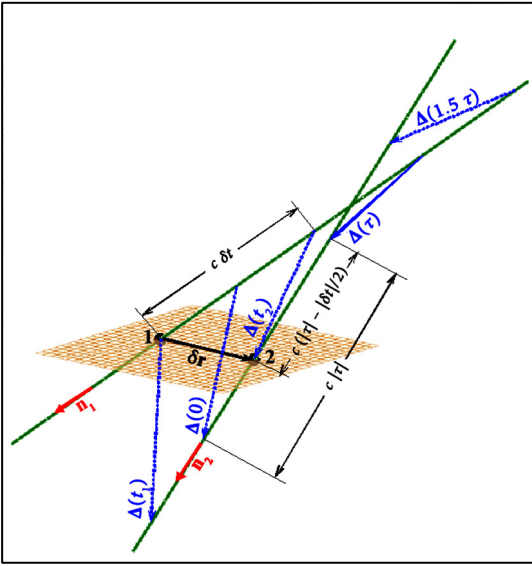

**Figure 1**. The scheme of two showers' representing points' motion. Two straight skew trajectories of the representing points are shown as green lines.
Symbols **1** and **2** indicate the shower cores' observation points. The vector δ**r** connects these points. The motion directing orts of both representing points are shown in red.
The vectors connecting the representing points at the specified moments are shown in blue. The special vector of the points' most close approach at the moment $\tau$ is shown too, as well as lengths of the characteristic segments of trajectories. The traversed path $c \cdot (|\tau| - |\delta t|/2)$ of the №2 representing point from the moment of the points' most close approach $\tau$ till the moment $t_2$ of the shower observation by the №2 goniometer is also shown.

At the moment $\tau$ of the points' closest approach the variable connecting vector obtains value:

$$\tilde{\Delta}(\tau) = \delta \mathbf{r} - \langle \mathbf{n} \rangle (c \, \delta t) - \frac{(\delta \mathbf{r}^T \delta \mathbf{n})}{(\delta \mathbf{n}^T \delta \mathbf{n})} \cdot \delta \mathbf{n},$$

so the respective relativistic invariant squared interval obtains the minimal value $\Delta^2 \equiv s^2(\tau) = \tilde{\Delta}(\tau)^T \cdot \tilde{\Delta}(\tau)$. The latter is independent of the choice of the inertial coordinate system such that the dispersion of its estimate can be determined in any inertial system, including the initial laboratory system. This well-defined quantity describes the estimation precision of nonvarying value $\Delta^2$, so it is invariant too. The dispersion $\sigma^2_{\Delta^2}$ of squared interval $\Delta^2$ as well as the dispersion $\sigma^2_{\tau}$ of the moment $\tau$ of the closest approach of the representing points can be estimated using the data from the table 2. The respective expressions are shown in the table 3.

We shall use the Lorentz-invariant space-like interval $\Delta = \sqrt{\Delta^2}$ between the representing points at the moment of their closest approach as a measure of the historical proximity of the showers of the given pair. The dispersion of this quantity is equal to $\sigma^2_{\Delta} = \sigma^2_{\Delta^2} / (4\Delta^2)$.

**TABLE 3   THE KINEMATIC CHARACTERISTICS OF TWO
SHOWER-REPRESENTING POINTS' APPROACH AND RESPECTIVE DISPERSIONS**

| System describing values | | Defining expressions |
|---|---|---|
| Coordinate time of the closest approach | | $\tau = -\dfrac{1}{c} \cdot \dfrac{(\delta \mathbf{r}^T \delta \mathbf{n})}{(\delta \mathbf{n}^T \delta \mathbf{n})}$ |
| Dispersion of time of the closest approach | | $\sigma_\tau^2 = \dfrac{1}{c^2 (\delta \mathbf{n}^T \delta \mathbf{n})^2} \left\{ \begin{array}{l} (\delta \mathbf{n}^T \mathbf{M} \delta \mathbf{n}) + (\delta \mathbf{n}^T \delta \mathbf{n})^2 (\delta \mathbf{r}^T \mathbf{D} \delta \mathbf{r}) + \\ +4(\delta \mathbf{r}^T \delta \mathbf{n}) \cdot \left[ (\delta \mathbf{r}^T \delta \mathbf{n})(\delta \mathbf{n}^T \mathbf{D} \delta \mathbf{n}) - (\delta \mathbf{n}^T \delta \mathbf{n})(\delta \mathbf{n}^T \mathbf{D} \delta \mathbf{r}) \right] \end{array} \right\}$ |
| Auxiliary quantities | Closest approach vector | $\boldsymbol{\Delta} = \delta \mathbf{r} - (c\, \delta t) \cdot \langle \mathbf{n} \rangle - \dfrac{(\delta \mathbf{r}^T \delta \mathbf{n})}{(\delta \mathbf{n}^T \delta \mathbf{n})} \cdot \delta \mathbf{n}$ |
| | Projection matrix $\mathbf{\Pi}$ | $\Pi^\mu_\nu = \left[ \delta^\mu_\nu \right] - \dfrac{\left[ \delta n_\nu\, \delta \mathrm{n}^\mu \right]}{(\delta n_\alpha\, \delta n^\alpha)}; \quad \mu, \nu, \ldots = x, y, z$ |
| | Special matrix $\mathbf{G}$ | $G^\mu_\nu = \left\{ \dfrac{\left[ \delta r_\nu\, \delta n^\mu \right]}{(\delta n_\alpha\, \delta n^\alpha)} + \dfrac{(\delta r_\beta\, \delta n^\beta)}{(\delta n_\alpha\, \delta n^\alpha)} \left[ \delta^\mu_\nu - 2 \dfrac{\left[ \delta n^\mu\, \delta n_\nu \right]}{(\delta n_\alpha\, \delta n^\alpha)} \right] \right\}$ |
| Squared interval of the closest approach | | $\Delta^2 = \boldsymbol{\Delta}^T \cdot \boldsymbol{\Delta}$ |
| Dispersion of the squared interval | | $\sigma_{\Delta^2}^2 = 4 (\boldsymbol{\Delta}^T \langle \mathbf{n} \rangle)^2 \cdot c^2 \sigma_{\delta t}^2 + \boldsymbol{\Delta}^T \left[ 4 (\mathbf{\Pi}^T \cdot \mathbf{M} \cdot \mathbf{\Pi} + \mathbf{G}^T \cdot \mathbf{D} \cdot \mathbf{G}) + (c\, \delta t)^2 \cdot \mathbf{D} \right] \boldsymbol{\Delta}$ |

\* The repeating indexes imply summation upon the spatial coordinates $x, y, z$.

*Application of the obtained quantities to the problem of identification
of shower couples with possibly congenetic ancestors*

The special pairs of showers, the ancestors of which may be suspected of having common origin, must obviously possess the following properties:

- The closest approach between the corresponding representing points takes place *before* the first observation of the shower by one of the goniometers, i.e. $\tau < \min(t_1, t_2) < 0$
- The space-like interval $\Delta$ between both representing points at the moment of their closest approach coincides with zero distance within the estimation precision, i.e. $\Delta < \sqrt{\sigma_\Delta^2}$.

Let us define the respective dimensionless verifying parameters distinguishing the desired couples of related showers among the whole set of the observed pairs.

It is convenient to match the time periods between the specific events in the given system of two showers' representing points with the period proper for this system, i.e. with the propagation time of electromagnetic pulse between the shower cores' observation points, i.e. $|\delta \mathbf{r}|/c$.

The earliest observation time of one of the shower cores $-|\delta t|/2$ is negative in accordance with the definition (2). Consequently, the time sequence order of two system events, exactly the earliest observation time of one of the shower cores and the moment of the closest approach of corresponding representing points, can be described by the dimensionless ratio:

$$\frac{\tau - \min(t_1, t_2)}{|\delta \mathbf{r}|/c} = \frac{c(\tau + |\delta t|/2)}{|\delta \mathbf{r}|} \tag{4}$$

This quantity is a ratio of two periods, related to the single physical system of the showers' pair, and measured in the fixed lab coordinate system. Both time periods are multiplied by the same gamma-factor under transformation to new inertial coordinate system, so the ratio (4) is the dimensionless relativistic invariant.

Unfortunately, the immediate use of the latter quantity is restricted by the technical difficulties arising from possibilities of large numerical values of either sign, which complicates pictorial presentation of all time sequencing possibilities. The usual approach of taking the logarithm of a large number can not be adopted in this case due to possible negative values; rather the well-known function of inverse hyperbolic sine [13] can be used instead. At large values of argument this function behaves asymptotically as the logarithmic one, preserving the argument sign, and is nearly linear for small values of argument. So let us define the *time **S**equencing* verifying dimensionless parameter $S$ for two events under consideration in the showers' system by the expression:

$$S = \operatorname{arsinh}\left(\frac{c(\tau + |\delta t|/2)}{|\delta \mathbf{r}|}\right); \qquad -\infty < S < +\infty.$$

The verifying parameter $S$ as well as the periods' ratio (4) takes on a negative value for those special shower pairs, for which the moment of closest approach of the representing points occurs *before* the earliest observation of either of the two showers. Only the shower pair with negative verifying parameter $S$ value can potentially be the historically proximal couple. The positive value of parameter $S$ means that the representing points approach to each other after the earliest observation of either of the two showers. (To be more exact, they would have a chance for mutual approach in future if they were not absorbed in the Earth under the goniometers.)

Let us describe the extent of the historic proximity of the representing points of the showers by the ratio $\Delta/\sqrt{\sigma_\Delta^2}$. If the ratio is less then unity (i.e. the invariant interval value not exceeds the limit of one standard deviation uncertainty), then the invariant distance between the representing points at the moment of the closest approach is effectively zero within the estimation precision. The shower pairs with the ratio not exceeding the unity can be considered as likely historically proximal couples. So long as this nonnegative ratio can attain large values let us define the *historical **P**roximity* verifying dimensionless parameter $P$ by the expression:

$$P = -\ln\left(\frac{\Delta}{\sqrt{\sigma_\Delta^2}}\right); \qquad -\infty < P < +\infty.$$

This parameter is relativistic invariant quantity as it is a function of two Lorentz-invariants. It is positive for the historically proximal showers, and the larger the value, the more reliable is the statement of the historical proximity of both showers.

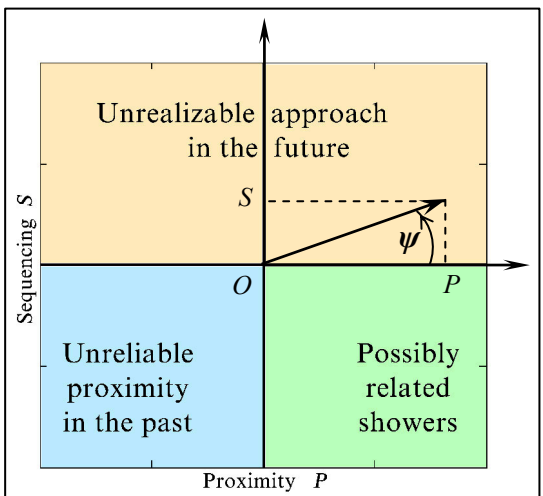

**Figure 2**. The plane of verifying invariant parameters for the pair of extensive air showers

The whole plane of two verifying dimensionless parameters $(P, S)$ can be divided into three zones as is shown in figure 2. The points in those zones are related:

1. With pairs of showers described by representing points which can approach one another in the future, but have to be absorbed in the ground: $S > 0$;
2. With pairs of showers described by representing points which have been in the near positions in their past, but the estimation precision of the approach invariant interval prevents the decision of true historical proximity: $S < 0, P < 0$;
3. With pairs of showers described by representing points which closely approach one another in their past and originate from a single point within the estimation precision: $S < 0, P > 0$.

Every point in the verifying parameters' plane defines the angle $\psi = \arctan(S/P)$ (see figure 2) between the *OP* axis and the direction from the origin *O* to the given $(P, S)$ point. With this angle we can define united verifying criterion via

$$K = \frac{2}{\pi} \cdot \text{arctg}\left(\frac{S}{P}\right); \qquad 0 \le K < 4 \,.$$

Depending on the value of criterion *K* the given $(P, S)$ point belongs to:
    $0 \le K < 2$ – Zone 1: unrealizable approach in the future;
    $2 \le K < 3$ – Zone 2: unreliable proximity in the past;
    $3 \le K < 4$ – Zone 3: possibly historically related showers.

So the united relativistic invariant criterion *K* allows unambiguous definition of the admissibility of assumption that the given pair of showers can have the historically proximal ancestors. The integer bounds of the possible *K*-zones are illustrative, delimiting the quadrants of the verifying parameters' plane.

### *The numerical check of the capability of the proposed verifying parameters and K-criterion*

For visual illustration of the verifying parameters capability the numerical simulation is performed of the flux of random shower pairs observed by two remote goniometers. The random generation uses simplified properties of showers as only the recognizing capability of *P*, *S* and *K* verifying values is under investigation, *rather than the physical properties* of true flux of the EAS.

The following conditions of the EAS observations are used:
- Both goniometers are identical.
- They are positioned on the ground at the same altitude and spaced by 100 *km* apart. The assumed vertical depth *q* of the atmosphere is a shower's 8 absorption ranges.
- It is assumed that the shower's core passes through the goniometer at the observation moment. The assumed standard deviation of the core coordinates is 200 *m* in the horizontal plane and zero in the vertical direction.
- The shower's observation timing is rounded within 1 *μs* period.
- The characteristic dispersion of the components estimation of the EAS velocity directional ort **n** is assumed equal to 0.001 for the showers from the zenith vicinity. The covariance matrix **D** of the (unite) ort is calculated assuming the dispersion dependence $\sim |\sec\theta|$ on the zenith angle $\theta$ of the shower's motion direction.

It is assumed for the toy simulation that *a half of the generated shower-pairs have nearly common origin*. The assumption, while *far from the physical reality,* allows an easy illustration of the recognizing capability of the proposed verifying parameters regarding the historical proximity of the showers' ancestors.

The unrelated showers in the pair are considered as the independent random phenomena with similar properties. They are generated under the following assumptions:
- Both goniometers have the same shower observation rate with the mean wait period equal to 10 *min*.
- There is no additional random shower observed by any of the goniometers in between detecting the two showers under consideration.
- The shower's motion direction azimuth is uniformly distributed along the horizon, while the distribution of the respective zenith angle of the shower's emergence point is used in a form of the flat atmosphere [14, 11] model: $\sim \sin(\theta)\cos(\theta)\exp(-q \cdot \sec(\theta))$ (see the q value definition)

These simple assumptions are adequate for the illustration purpose.

The kinematic characteristics of the premeditatedly historically proximal showers in the pair are generated with use of next schematic model:

- The random ort in the direction of the ancestors' emergence region is estimated quite as the reversed directional ort of the single independent shower propagation.
- It is assumed that the pair appears at a distance of more than 500 *km* from the middle point between the goniometers.
- It is assumed that the distribution of the additional distance *R* above the mentioned minimal distance to the appearance point has form $\sim R^3 \cdot \exp(-R/L)$; here the length parameter $L = 10^5 \, km$ describes decrease of generation points' density with the distance from the Earth. The adopted distribution is only a reasonable assumption.
- The emergence points of both ancestors of the creating showers for the future pair are slightly moved apart from the common point defined above to simulate the random amount of their proximity measure.
- The necessary values of the observation times and motion orts are calculated via the equations of uniform motion of both representing points by use of already known coordinates of the emergence points and observation points with the assumption of simultaneous emergence of both ancestors.

The data generated in accordance with the above scheme is sufficient for estimation of the verifying parameters for every pair of showers. The data are simulated for $10^6$ shower pairs.

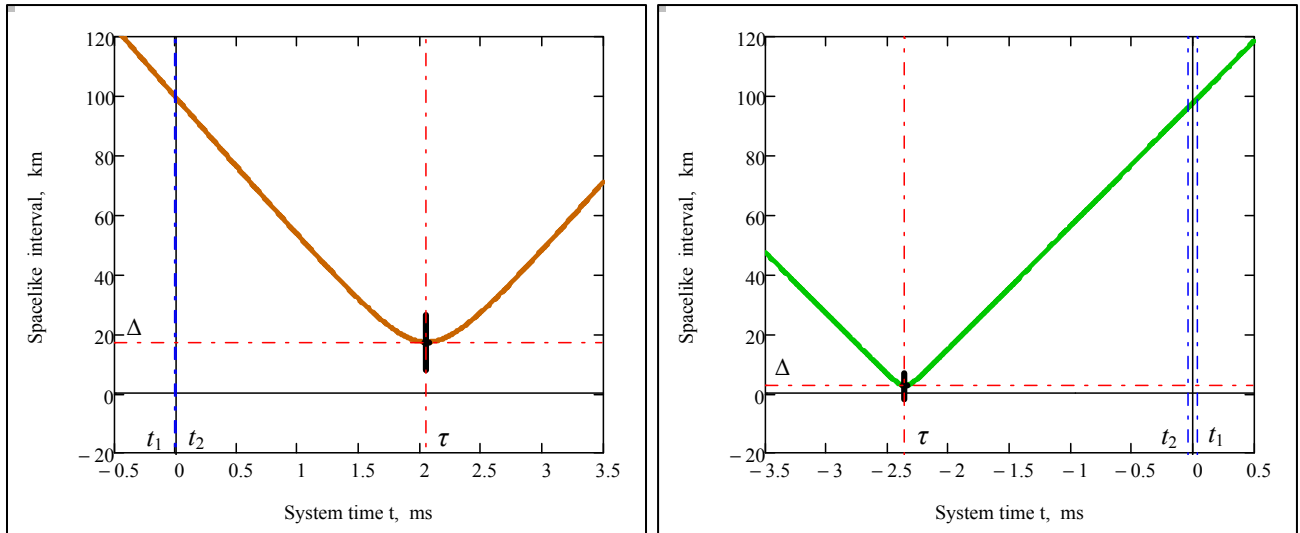

**Figure 3.** Two selected examples of the time dependencies of the space-like intervals between the representing points of the showers.

The brown curve on the left side displays the interval dependence for the first pair of showers when representing points can probably approach each other *after* the (earliest) observation of the shower №1.
The calculated kinematic characteristics, verifying parameters and criterion values for the first pair of showers are:
$t_1 = -3.5 \cdot 10^{-3} \, ms$, $t_2 = 3.5 \cdot 10^{-3} \, ms$, $\tau = (2.06 \pm 0.02) \, ms$; $\Delta = (17.2 \pm 9.2) \, km$; $P = -0.63$, $S = +2.52$, $K = 1.16$.

The green curve on the right side displays the similar dependence in the case when representing points approach have take place *before* the (earliest) observation of the shower №2. In this second example the interval of the closest approach is indistinguishable from the single common point of two showers' ancestors' appearance.
The calculated kinematic characteristics, verifying parameters and criterion values for the second pair of showers are:
$t_1 = 0.039 \, ms$, $t_2 = -0.039 \, ms$, $\tau = (-2.36 \pm 0.02) \, ms$; $\Delta = (2.6 \pm 4.3) \, km$; $P = +0.52$, $S = -2.64$, $K = 3.13$.

There are shown in black the 1σ-spans of the estimations of the closest approximation space-like intervals.

Two cases of the dependences of intervals $|\tilde{\Delta}(t)|$ (3) between two representing points on the system evolution time are shown in the figure 3. These illustrative examples are generated under above described assumptions. The brown curve on the left side displays the interval dependence for the shower pair with representing points approaching each other *after* the earliest observation of the one of the showers. The green curve on the right side displays the similar dependence in the case when representing points' closest approach takes place *before* the earliest observation of one of the showers. In the latter case the interval of the closest approach is indistinguishable from the single common point of the two showers' ancestors' emergence.

The visual positions of the estimated characteristic points ($P$, $S$) in the verifying parameters' plane (compare with the figure 2) are shown in the figure 4. They form two almost separate clusters: the brown points in the upper half plane represent the pairs of showers which correspond to unrealizable proximity in their future, while the points in the lower half-plane represent the pairs of realized proximity in their past. The green points in the right side correspond to the pairs with comparatively reliable estimation of their proximity and the blue points in the left side correspond to the uncertain proximity.

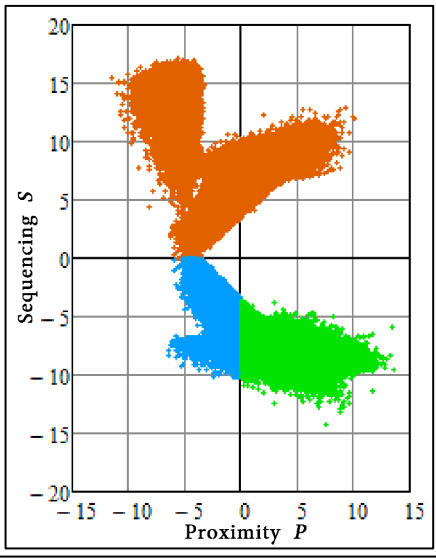

**Figure 4.** The verifying parameters plane populated by the points corresponding to the shower-pairs in toy simulation

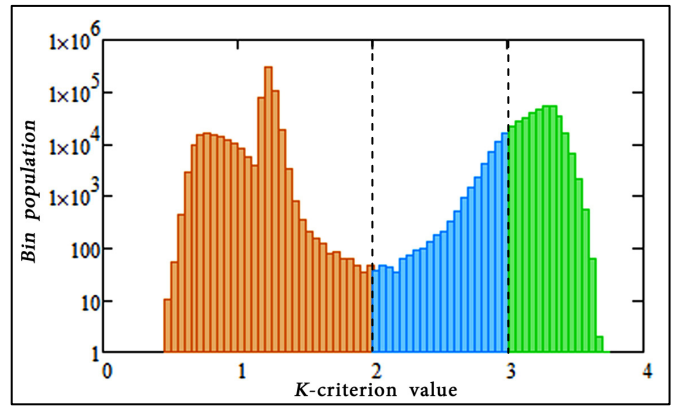

**Figure 5.** The $K$-criterion histogram is divided in compliance with the three possible types of proximity of the shower-pairs from the same sample. The segment bounds of the three possible zones of $K$ values are shown too.

Distribution of $K$-criterion variable for all three possible types of proximity is shown in figure 5 using the same color-codes. As discussed earlier, there are three distinct ranges for $K$-criterion value:
1. $0 \leq K < 2$ (the histogram shown in brown): the showers' representing points can approach one another in the *future*, but have to be absorbed in the ground.
2. $2 \leq K < 3$ (the histogram shown in blue): the showers' representing points have been in the closest positions in their *past*, but the standard deviation segment of the approach interval estimation *do not cover* the $\Delta$-interval value, so the showers' historical proximity is questionable.
3. $3 \leq K < 4$ (the histogram shown in green): the showers' representing points have, possibly, closely approached one another *before* the observations of the corresponding showers and the estimation of the $\Delta$-interval of their closest proximity *is covered* by the relevant standard deviation; the showers of those pairs can be congeneric.

It should be stressed again that the density of points in the ($P$, $S$) plane in figure 4 as well as the bin population in the histogram in figure 5 *has NO connection with physical problem* of observations of historically proximal couples of showers; rather they served useful illustrative purpose.

*Conclusion*

The paper introduces relativistically invariant verifying parameters ($P$, $S$) as well as combined variable $K$ which can be used for the preliminary classification of the shower pairs observed by the remote EAS-studying installations as possibly originating from single process in space. The numerical example to illustrate the discriminative power of these variables is also presented.

It is clearly shown that once the shower couples with the $K$-criterion values in the range $3 \le K < 4$ are detected they must be subject to individual thorough investigation. The shower couples with the criterion in the range $2 \le K < 3$ can be studied provisionally as long as the verification parameters' estimations depend on the precision of measurements accomplished by both goniometers. At the very least, kinematic parameters, i.e. (see table 3) the period $\tau$ up to the closest approach time and the value of the space-like interval $\Delta$ between the ancestors of showers of the couple at their closest approach moment, must be examined. There are precisely the estimators (together with their dispersions) that allow assumption of the real physical connection among the ancestors of both showers in their past. The more in-depth analyses of the properties of the showers' ancestors' is feasible only if some additional data on both showers are available.

*Acknowledgements*

The authors are grateful to our current and former colleagues for their technical support and sympathetic attention to our investigations and respective results. We are especially thankful to our colleagues working now abroad.

*References*

1. Warsaw Workshop on Non-Standard Dark Matter (2016):
   http://indico.fuw.edu.pl/conferenceDisplay.py?confId=45

2. Ochi N., Iyono A., Kimura H. et al. "Search for large-scale coincidences in network observation of cosmic ray air showers". J. Phys G: Nuclear and Particle Physics (2003) **29**, №6, pp.1169-1180.

3. Atsushi Iyono, Hiroki Matsumoto, Kazuhide Okei et al:
   "Parallel and Simultaneous EAS events due to Gerasimova-Zatsepin effects observed by LAAS experiments"; proceedings of the 31st ICRC (2009), Łódź;
   http://icrc2009.uni.lodz.pl/proc/pdf/icrc0941.pdf

4. Karel Smolek, Filip Blaschke, Jakub Čermak et al: "ALTA/CZELTA – a sparse very large air shower array: overview of the experiment and first results"; proceedings of the 31st ICRC (2009), Łódź;
   http://icrc2009.uni.lodz.pl/proc/pdf/icrc1300.pdf

5. F. Blaschke, J. Čermák, J. Hubík et al: "CZELTA: An overview of the CZECH large-area time coincidence array"; Astrophys. Space Sci. Trans. (2011), **7**, pp.69–73;
   www.astrophys-space-sci-trans.net/7/69/2011

6. Yu. Verbetsky, M. Svanidze, A. Iashvili, E. Tskhadadze and D. Kokorashvili:
   "First results on the spatiotemporal correlations of the remote Extensive Air Shower pairs". 23rd ECRS (and 32nd RCRC) Moscow, J. Phys.: Conf. Ser. 409 (2013) 012085;
   http://iopscience.iop.org/article/10.1088/1742-6596/409/1/012085/meta;jsessionid=89322C00C25D03FA0B0D63C56F0B2333.c5.iopscience.cld.iop.org

7. M. Abbrescia, L. Baldini, R. Baldini Ferroli et al.: "Search for long distance correlations between extensive air showers detected by the EEE network", Eur. Phys. J. Plus (2018) **133** p.34

8. Cosmic-Ray Extremely Distributed Observatory (CREDO): http://credo.science/

9. Yuri Verbetsky, Manana Svanidze, Abesalom Iashvili, Levan Kakabadze:
   "Extensive Air Showers' Arrival Direction Distribution by TBS Array",
   International Journal of High Energy Physics.(2014) **1**, №4, pp.49-54,
   http://www.sciencepublishinggroup.com/journal/archive.aspx?journalid=124&issueid=-1

10. Manana Svanidze, Yuri Verbetsky, Ia Iashvili, et al: "Angular distribution of extensive air showers by TEL array under GELATICA experiment"; GESJ: Physics (2016) №1(15) pp.54-62;
http://gesj.internet-academy.org.ge/download.php?id=2740.pdf

11. Yu. G. Verbetsky, M. S. Svanidze, A. Iashvili, I. Iashvili, L. Kakabadze:
"Extensive air showers' arrival direction distribution by the TSU array under GELATICA experiment"; GESJ: Physics (2018) №2(20) pp.43–55;
http://gesj.internet-academy.org.ge/download.php?id=3175.pdf

12. Yuri Verbetsky, Manana Svanidze, Abesalom Iashvili, Ia Iashvili, Levan Kakabadze, Nino Jonjoladze: "Investigation of the arrivals' directions differences for consecutive Extensive Air Showers using the data taken by TEL goniometer under GELATICA network", GESJ: Physics (2019) №2(22) pp.27–33;
http://gesj.internet-academy.org.ge/download.php?id=3274.pdf

13. Handbook of mathematical functions.
Under edition: Abramowitz M. and Stegun I. A. National bureau of standards, (1964)

14. Sokolsky P.: "Introduction to Ultrahigh Energy Cosmic Ray Physics" p.200; Boulder (Colorado, USA): Westview Press (2004).